# Terahertz-Frequency Plasmonic-Crystal Instability in Field-Effect Transistors with Asymmetric Gate Arrays

G. R. Aizin[1,*], S. Mundaganur[2], A. Mundaganur[2], and J. P. Bird[2]

1: Kingsborough College, The City University of New York, NY 11235, USA

2: Department of Electrical Engineering, University at Buffalo,

The State University of New York, Buffalo, NY 14260

We present a theory for plasmonic crystal instability in a semiconductor field-effect transistor with a dual grating gate, designed with strong asymmetry in the crystal elementary cell. Under the action of a dc current bias, we demonstrate that Bloch plasma waves in the resulting plasmonic crystal, formed in this transistor, develop the Dyakonov-Shur instability across the entire Brillouin zone. By calculating the energy spectrum of the plasmonic crystal and its instability increments, we analyze the dependence of the latter on the electron drift velocity and the extent of the structural asymmetry. Our results point to the possibility of exciting radiating steady-state plasma oscillations at room temperature, in transistors with asymmetric gate arrays that should be readily implementable via standard nanofabrication techniques. Long-range coherence of the unstable plasma oscillations, generated in the elementary cells of the crystal, should dramatically increase the radiated THz electromagnetic power, making this approach a promising pathway to the generation of THz signals.

* Gregory.Aizin@kbcc.cuny.edu

# I. INTRODUCTION

The design of efficient tunable on-chip sources of terahertz (THz) electromagnetic (EM) radiation remains a challenging technological problem. Such sources are one of the key required elements of nanoscale, beyond-6G, wireless communications systems, with potential applications ranging from wireless networks-on-chip [1] to biomedical systems [2]. All-electronic on-chip THz sources include Schottky diode frequency multipliers [3], GaN IMPATT (impact ionization avalanche transit-time diode) diodes [4], and resonant tunneling diodes [5], all of which operate at the lower end of the THz spectrum (~0.1 -1 THz) [6], and plasmonic THz field-effect transistors (TeraFETs) with the capability to operate at frequencies up to 10 THz [7].

Plasmonic TeraFETs use the instability of collective plasma oscillations in the two-dimensional (2D) electron channel of a transistor to generate EM radiation. The channel (or some portion of it covered by a gate) functions as a plasmonic cavity, reflecting plasma waves at its opposite ends. Dependent upon the specific reflection conditions at these boundaries, interference of plasma waves travelling back and forth can lead to a buildup of the plasma-wave amplitude in the cavity. As first predicted by Dyakonov and Shur [8], when the reflection conditions differ substantially, a constant (dc) drive current may induce plasma instability, resulting in an exponential time-dependent growth of the plasma-wave amplitude in the transistor. In the resultant steady-state, this leads to EM radiation at the plasma frequencies [9,10]. Since these frequencies lie in the THz range, and are easily tunable via the gate voltage of the transistor, plasmonic TeraFETs are potentially attractive for a variety of applications.

Previous experimental studies have demonstrated signatures of the Dyakonov-Shur (DS) instability in transistor-based plasmonic cavities and have also revealed the presence of THz EM radiation [11,12]. Recently, transistors with specially engineered structural asymmetry have been implemented, and features characteristic of the DS instability were found in their transistor characteristics at temperatures as high as 350 K [13]. In spite of this, the radiated EM power measured in previous experiments [11,12] has been too small (~nW) to enable practical applications. To improve the emitted power of such radiation transistors can be combined into arrays [14], but the lack of coherence among plasma oscillations, generated in the individual devices, impedes any significant power increase.

The most promising approach to increase the radiated power from plasmonic sources is to design transistors in which a suitable periodic modulation of the electron channel is introduced. This idea has long motivated the study of transistors with "grating-gate" geometries [15] and has more recently been proposed in the context of a plasmonic transistor with periodically modulated channel width [16]. Regardless of the specific manner in which it is implemented, periodic modulation of the 2D electron channel leads to the formation of a plasmonic crystal, with an energy spectrum that was first described theoretically in [17] and confirmed experimentally in [18,19]. In such a system, plasma oscillations in the individual elementary cells are coupled to one another via EM interaction, and the coherence of these oscillations results in the formation of a Bloch plasma wave across the entire crystal. Under such conditions, once plasma waves become unstable in the individual elementary cells, a drastic enhancement of the radiated EM power should occur due to coherent addition of the EM waves generated in different cells [16,20,21].

In spite of the many studies that have addressed the issue of plasmon generation in transistors incorporating different gate-array geometries [15-24], the need for practical schemes for on-chip THz-signal generation nonetheless remains. Here, we therefore present a quantitative theoretical study of plasma-wave generation in the channel of a high electron mobility transistor with a custom grating-gate design. This array structure is designed to achieve strong asymmetry within the elementary cells of the resulting plasmonic crystal. Through a hydrodynamic treatment of the electron fluid within the FET, we demonstrate that a dc bias current may cause the plasma band modes of this periodic system to become unstable within the *entire* Brillouin zone. (This situation should be contrasted with that in plasmonic crystals with *symmetric* reflection conditions, in which the instability may develop in only a *few limited ranges* of the Bloch wavevector [20].) In this way, we identify the conditions under which the plasma-wave instability increment (*i.e.*, the growth rate of the plasma wave amplitude when it reflects back and forth within its unit cell [8]) can exceed the damping rate of the plasma waves due to (phonon and impurity) scattering.

# II. THEORY OF A PLASMONIC CRYSTAL IN AN ASYMMETRICALLY GATED TRANSISTOR

The transistor design that we propose is shown schematically in Fig. 1(a). The channel of this device consists of periodically spaced, gated and ungated, sections, of respective lengths $L_1$ and $L_2$. Each cell of the array comprises a pair of dissimilar gates, the wider of which is biased to generate the periodically modulated density profile shown in Fig. 1(b). In the discussion that follows, we denote the equilibrium electron density in the gated and ungated sections as $n_{01}$ and $n_{02}$, respectively ($n_{01} < n_{02}$). The narrow gate within each cell is considered to be grounded, with its purpose being to produce an additional capacitive link [24] between the 2D electron channel and one of the edges of the wider gate (the left edge as shown in Fig. 1(a)). This narrow element is therefore the source of the asymmetry that is crucially required to achieve plasma-wave instability in the plasmonic cavities formed under the wide gates. In Fig. 1(c) we demonstrate the feasibility of realizing the periodic gate geometry via nanolithography.

When electron-electron interactions are the dominant source of scattering within the system, plasmonic excitations in the gated sections of the 2D electron channel can be described in the hydrodynamic approximation. Neglecting the influence of random electron scattering (*i.e.*, assuming that $\omega \tau \gg 1$ where $\omega$ is the plasma frequency and $\tau$ is momentum relaxation time), the relevant hydrodynamic equations for the local electron-fluid density ($n(x,t)$) and velocity ($v(x,t)$) are the continuity equation and the Euler equation:

$$\frac{\partial n}{\partial t} + \frac{\partial (nv)}{\partial x} = 0, \quad \frac{\partial v}{\partial t} + v \frac{\partial v}{\partial x} = \frac{e}{m^*} \frac{\partial \varphi}{\partial x}. \tag{1}$$

Here, $\varphi(x,t)$ is the electric potential in the channel, and $-e$ and $m^*$ are the electron charge and effective mass, respectively.

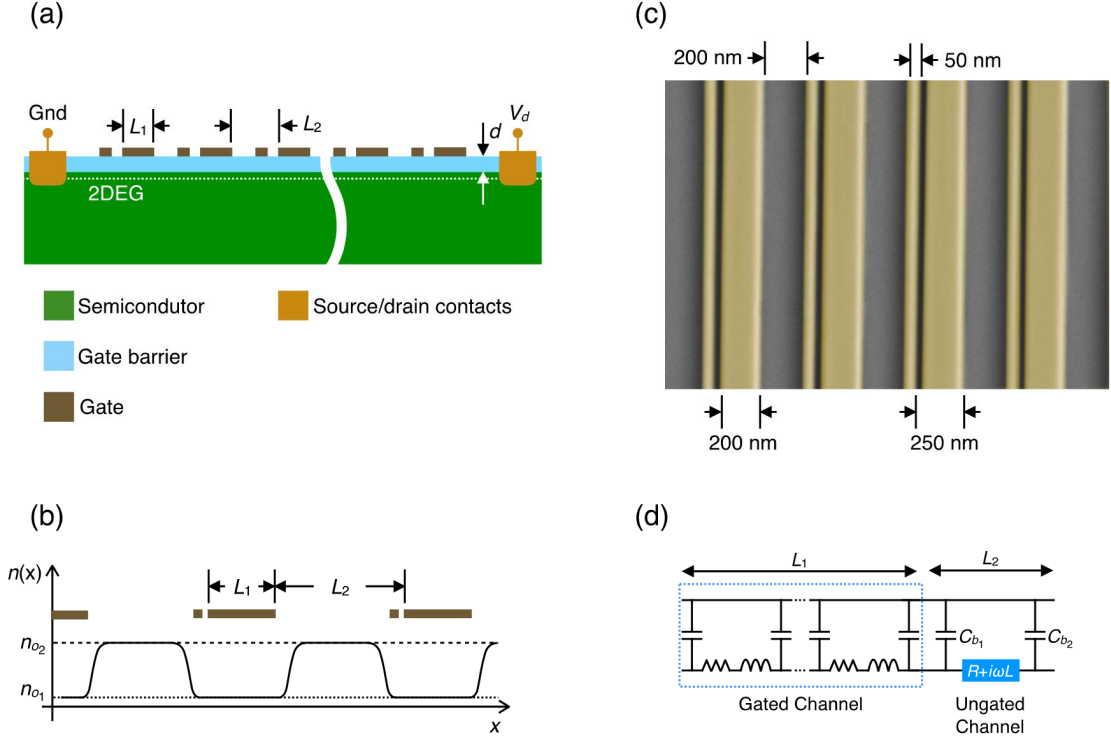

**Fig. 1. Double Column.** (a) Cross sectional schematic showing the basic structure of the suggested grating-gate device. (b) Density profile induced in the device by biasing the wide gates. (c) False-color electron micrograph showing a grating-gate fabricated on a GaAs substrate. Metal gates are Cr/Au (5-/50 nm). (d) Equivalent electric circuit diagram of the device of panels (a) – (c). See text for further details.

To solve Eq. (1), we linearize it assuming small fluctuations of the electron density ($n = n_{01} + \delta n$) and velocity ($v = v_0 + \delta v$, where $v_0$ is the electron drift velocity in the plasmonic cavity with constant source-drain current). In the gated sections of the channel, fluctuations of the electron density and the electric potential ($\delta\varphi$) are connected via $-e\delta n = C\delta\varphi$, where $C = \varepsilon\varepsilon_0/d$ is the capacitance per unit area between the channel and gate, and $d$ and $\varepsilon$ are the thickness and dielectric constant, respectively, of the gate barrier (see Fig. 1(a)). In this quasistatic limit, solution of the linearized form of Eq. (1) can be written for the Fourier harmonics $\delta n_\omega$ and $\delta v_\omega$, both proportional to $e^{-iqx+i\omega t}$, as [25]:

$$I_\omega(x,t) = \left(I_1 e^{-iq_1 x} + I_2 e^{-iq_2 x}\right) e^{i\omega t},$$

$$V_\omega(x,t) = \frac{1}{CW}\left(\frac{I_1}{v_0+v_p}e^{-iq_1 x} + \frac{I_2}{v_0-v_p}e^{-iq_2 x}\right) e^{i\omega t}. \qquad (2)$$

In these expressions, $I_\omega(x,t) = -e(v_0 \delta n_\omega + n_0 \delta v_\omega)W$ is the plasmonic current in a channel of width $W$, while $V_\omega(x,t) = \delta\varphi_\omega$ is the voltage distribution within the plasma wave. The wavevectors $q_{1,2} = \omega/(v_0 \pm v_p)$ describe acoustic plasma waves propagating with ($q_1$) and opposite to ($q_2$) the electron drift, at respective velocities $v_0 + v_p$ and $v_0 - v_p$. $v_p = \sqrt{e^2 n_{01}/m^* C}$ is the velocity of the acoustic plasmons in the gated 2D channel in the absence of drift [26] and the constant coefficients $I_1$ and $I_2$ are determined by the boundary conditions. The gated sections of the transistor channel serve as a waveguide for the plasma waves and can be represented by an equivalent-circuit diagram (Fig. 1(d)) comprising a transmission line (TL) with distributed capacitance $\mathcal{C}_1 = CW$, kinetic inductance $\mathcal{L}_1 = m^*/e^2 n_{01} W$, and (if scattering is included) resistance $\mathcal{R}_1 = \mathcal{L}_1/\tau$ (all defined per unit channel length) [27,28].

In the ungated sections of the transistor channel (where $n_{02} \gg n_{01}$) the plasma eigenfrequencies are well separated from those in the gated sections (if both sections are of comparable length). Consequently, we are able to restrict our analysis to frequencies close to those of the gated plasmons. In this limit, plasma oscillations in the ungated regions of the channel, having much higher frequency, are not excited. In our equivalent-circuit diagram (Fig. 1(d)), we therefore represent the ungated sections of the channel by a lumped inductance $\mathcal{L}_2 = m^* L_2/e^2 n_{02} W$ and resistance $\mathcal{R}_2 = \mathcal{L}_2/\tau$, yielding total impedance $Z_u = \mathcal{R}_2 + i\omega\mathcal{L}_2$. In the limit $\omega\tau \gg 1$, we use $Z_u \approx i\omega\mathcal{L}_2$ in our calculations.

The boundary between gated and ungated sections of the channel represents a region of special interest. As shown in [24], a fringing capacitance ($C_b$, see Fig. 1(d)) is formed between the 2D electron layer and the gate edge; this capacitance serves as a shunt, allowing some part of the plasmonic current to flow through the gate, while simultaneously leading to a step-like change in that flowing through the channel. The value of the fringing capacitance can be estimated as $C_b = \alpha\varepsilon\varepsilon_0 W$, where $\alpha = (1/\pi)\ln(\pi L_2/d) \sim 1$ is a geometric factor [24]. We also assume continuity of the electric potential across the boundary between the gated and ungated regions, something that is justified if the length of the transition region is smaller than the electron-electron scattering length (*i.e.*, if ballistic transport across this boundary is assumed [20]). Due to the deliberate asymmetry that we create by placing an additional narrow metal finger on one side (the left) of the wider gate (see Figs. 1(a) & 1(c)), the fringing capacitances will be different at the opposite edges of this gate (*i.e.*, $C_{b_1} \neq C_{b_2}$).

The generic plasma dispersion equation for the 1D plasmonic crystal can be written using the Bloch theorem and the transfer matrix $\hat{T}$ connecting the values of $I_\omega$ and $V_\omega$ at opposite sides of the crystal elementary cell [16]:

$$det\hat{T} - e^{ikL}Tr\hat{T} + e^{2ikL} = 0, \qquad (3)$$

where $k \in [-\pi/L, \pi/L]$ is the Bloch wavevector and $L = L_1 + L_2$ is the length of the elementary cell (see Fig. 1(a)). The transfer matrix for the gated section ($\hat{t}_g$) found from Eq. (2) is:

$$\hat{t}_g = e^{-iM\theta} \begin{pmatrix} \cos\theta - iM\sin\theta & iZ_0 \sin\theta \\ \frac{i}{Z_0}(1-M^2)\sin\theta & \cos\theta + iM\sin\theta \end{pmatrix}, \qquad (4)$$

where $\theta = \omega L_1/(v_p(1-M^2))$, $Z_0 = 1/CWv_p$ is the characteristic impedance of the plasmonic TL [28] and $M = v_0/v_p$ is the Mach number in the electron fluid. Transfer matrices for the ungated section ($\hat{t}_u$) and the boundary between gated and ungated sections ($\hat{t}_b$) are:

$$\hat{t}_u = \begin{pmatrix} 1 & Z_u \\ 0 & 1 \end{pmatrix}, \quad \hat{t}_b(Z_{b1,2}) = \begin{pmatrix} 1 & 0 \\ \frac{1}{Z_{b1,2}} & 1 \end{pmatrix}, \qquad (5)$$

where $Z_{b1,2} = 1/i\omega C_{b1,2}$. The transfer matrix $\hat{T}$ can be written in terms of the matrices in Eqs. (4) and (5):

$$\hat{T} = \hat{t}_g \hat{t}_b(Z_{b1}) \hat{t}_u \hat{t}_b(Z_{b2}). \qquad (6)$$

By substituting Eq. (6) into Eq. (3) we obtain the plasma dispersion equation:

$$\cos(kL + M\theta) = \left[1 + \frac{Z_u}{2}\left(\frac{1}{Z_{b1}} + \frac{1}{Z_{b2}}\right)\right]\cos\theta$$
$$+ \frac{i}{2}\left[Z_0\left(\frac{1}{Z_{b1}} + \frac{1}{Z_{b2}}\right) + \frac{Z_0 Z_u}{Z_{b1}Z_{b2}} + MZ_u\left(\frac{1}{Z_{b1}} - \frac{1}{Z_{b2}}\right) + (1 - M^2)\frac{Z_u}{Z_0}\right]\sin\theta. \quad (7)$$

In the next section this equation is solved numerically, allowing the conditions for plasma-wave instability to be found.

### III. RESULTS

We have solved Eq. (7) numerically and found complex plasma frequency ($\omega = \omega' + i\omega''$) as a function of the Bloch vector $k$. The real part of $\omega$ corresponds to the frequency ($\omega'/2\pi$) of the plasma wave. The imaginary part, on the other hand, determines the instability increment ($\omega'' < 0$) or decrement ($\omega'' > 0$), that is the growth or decay of the plasma wave, respectively, as it oscillates underneath the gate. In Fig. 2, we plot the calculated dispersions $\omega'(k)$ and $\omega''(k)$, for the first five plasmonic bands and for four different values of the Mach number (panels (a) – (d)). Motivated by experiments such as that of [13], we consider an InGaAs-based TeraFET ($m^* = 0.042 m_o$, where $m_o$ is the free-electron mass; $\varepsilon = 13$) with $L_1 = L_2 = 200$ nm, $n_{01} = 3 \times 10^{11}$ cm$^{-2}$, $n_{02}/n_{01}=3$, and $d/L = 0.05$. We also assume that the narrow metal finger positioned on the left side of each wide gate increases the fringing capacitance so that $C_{b_2}/C_{b_1} \equiv \gamma > 1$. The parameter $\gamma$ therefore characterizes the asymmetry of the boundary conditions at the opposite edges of each of the gate-defined plasmonic cavities. In the calculations here we assume $\gamma = 2$, a reasonable value that should be readily attainable in experiment.

In Fig. 2(a), we show the real (upper panel) and imaginary (lower panel) components of $\omega$ when no dc drift is present in the system (*i.e.*, $M = 0$). The figure plots the first five plasmonic bands ($N = 0 - 4$) generated by the coupling among the different cells, with $N = 0$ corresponding to the acoustic mode of the plasmonic crystal. The lower panel shows that $\omega'' = 0$ for all of the modes, a general result when $M = 0$. In other words, no instability or additional damping of the plasma waves occurs for this condition. The situation changes dramatically when a dc current is passed through the transistor (corresponding to the Mach numbers $M > 0$, panels (b) – (d)). All modes now develop a non-zero imaginary component of their plasma frequency, with $\omega'' < 0$ for all modes except for the acoustic one (and partially for the $N = 1$ mode). The negative sign of $\omega''$ corresponds to instability and to exponential growth ($\sim e^{|\omega''|t}$) of the plasma waves with increment $|\omega''|$. The growth occurs due to the dissimilar reflection experienced by the drifting plasma waves at the asymmetric boundaries of each gated cavity. The asymmetry of these boundaries (and of the reflections that they generate) finally results in a net transfer of the kinetic energy carried by the drifting plasmon into radiated EM energy. While this situation holds when the drift is from the boundary with low terminal impedance to that with high impedance (*i.e.*, when $M > 0$), reversing the direction of the drift results in attenuation of the plasma waves [8]. This follows straightforwardly from the form of Eq. (7), in which drift-velocity reversal ($M < 0$) is equivalent to changing the solution for the plasma frequency to its complex conjugate $\omega^*$, such that the plasma-wave increment ($\omega'' < 0$) is converted to a decrement ($\omega'' > 0$).

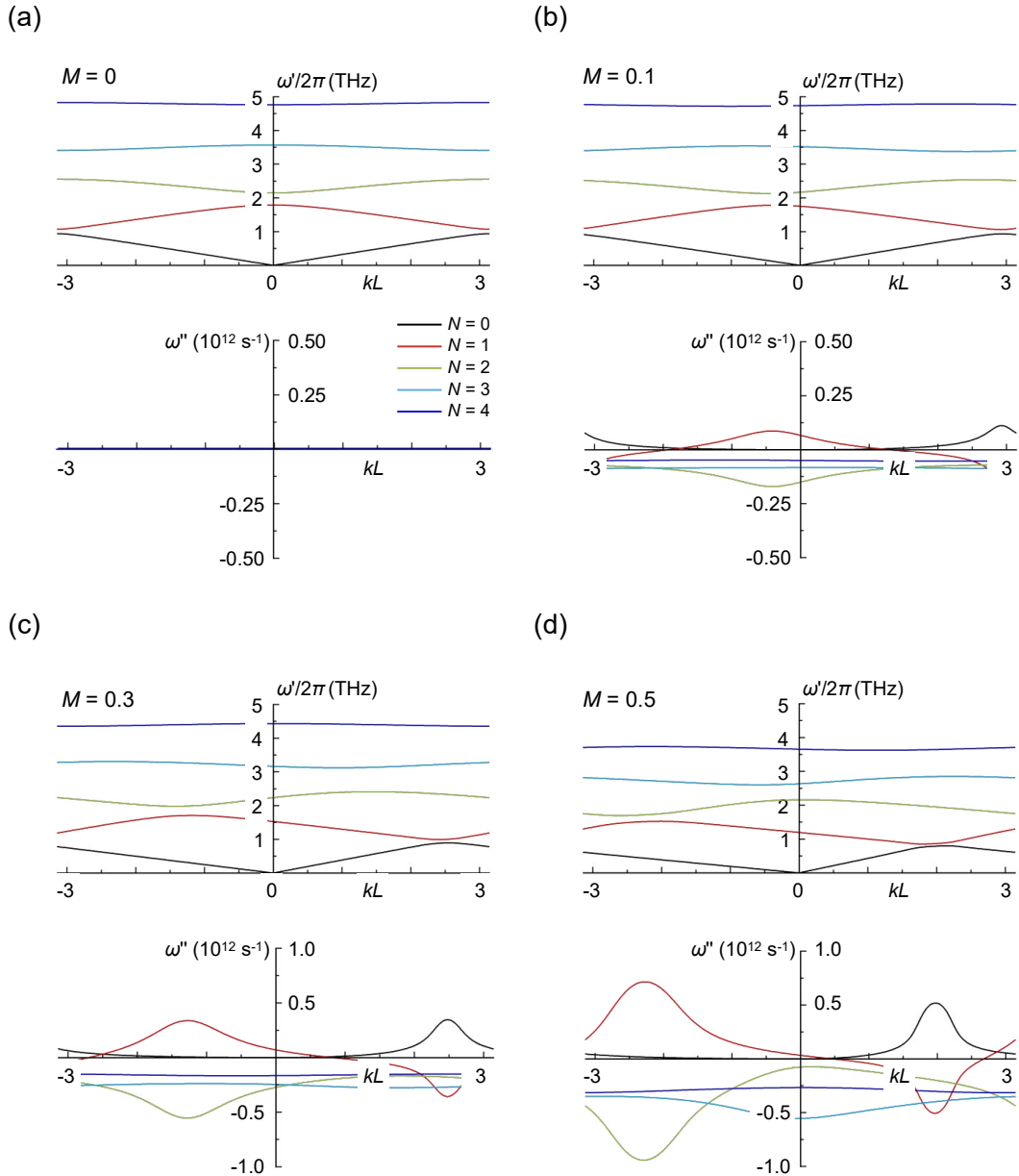

**Fig. 2. Double Column.** Variation of real (upper panel) and imaginary (lower panel) components of the plasma frequency for various values of the Mach number ($M$) and for $N = 0, 1, 2, 3$ & 4 (see lower panel of (a) for legend defining correspondence to lines colors). (a) $M = 0$. (b) $M = 0.1$. (c) $M = 0.3$. (d) $M = 0.5$. Here, $N$ is the plasmonic band index with $N = 0$ corresponding to the acoustic plasmonic crystal mode.

If we consider how the behavior of the system evolves with increasing dc bias, corresponding to increasing Mach number, Figs. 2(b) – 2(d) show that $\omega''$ systematically becomes more negative as $M$ is increased to 0.5, corresponding to strengthening instability. Although not included here, our numerical analysis shows that $\omega''$ reaches its most negative values at $M \sim 0.9$, before dropping back towards zero at $M \to 1$. In the results presented in Fig. 2, we restrict our analysis to $M \leq 0.5$, reflecting the fact that the maximum drift achievable in practice is limited by velocity saturation.

It is evident from the results of Fig. 2 that $\omega''$ varies in complicated fashion across the Brillouin zone, in a manner strongly dependent upon the band index $N$, a consequence of the changing symmetry of the coupled (quantized) modes in gate-defined cavities. In Fig. 2, $\omega'' < 0$ in the entire Brillouin zone for all bands with $N \geq 2$. The instability increment $|\omega''|$ is maximal for $N = 2$ and it is this mode that will therefore dominate the plasmonic response of the transistor. From inspection of the upper panel of Fig. 2(d), we see that for this band the real part of $\omega$ corresponds to a plasma frequency of around 2 THz, pointing to the promise of this system for the generation of THz signals. While the instability should occur for any value of $k$ within the Brillouin zone, it is nonetheless strongest at some particular wavevector (whose value depends upon the Mach number). Since the Bloch vector determines the phase difference between oscillations in different elementary cells of the crystal, this result indicates the importance of phase matching of the unstable plasma oscillations in the different plasmonic cavities [20].

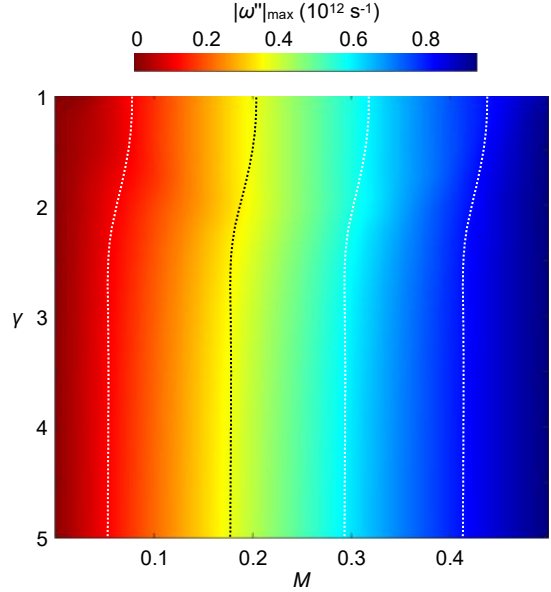

**Fig. 3. Single column.** Two-dimensional contour showing the variation of the maximum plasma-wave increment $|\omega''|_{max}$ as a function of $\gamma$ and $M$. Dotted contour lines serve as a guide to the eye to exhibit the dependence of $|\omega''|_{max}$ on $\gamma$.

The plasma-wave increment also depends on the asymmetry factor ($\gamma$) of the array, as we explore in Fig. 3. Here we show the variation of the maximum value of the instability increment $|\omega''|_{max}$ as a function of $\gamma$ and $M$. The contour indicates that $|\omega''|_{max}$ increases as $\gamma$ is increased from 1 to around 3, following which it appears to saturate. This suggests that the influence of the asymmetric boundary conditions on the instability increment essentially reaches the limiting value, expected for ideally asymmetric boundaries (*i.e.*, for $\gamma = \infty$ [8]) once $\gamma = 3$. This is a very encouraging result, as an asymmetry factor of $\gamma = 3$ should be readily attainable by established nanolithography techniques.

Although not shown here, it should also be pointed out that the instability persists in the symmetric limit at $\gamma = 1$ once $M \gtrsim 0.1$. However, in this case the instability only develops over some narrow ranges of the Bloch wave vector and therefore requires very precise phase-matching conditions between the elementary cells. The plasma instability in symmetric arrays was first considered in Ref. [20], and our results at $\gamma = 1$ are consistent with the findings of this paper. The structural asymmetry that we observe here, on the other hand, triggers the DS instability in the unit cells of the array. Since this occurs across the entire Brillouin zone (see Fig. 2) it should render demonstration of the instability easier in practice, where perfect gate periodicity is difficult to maintain. In the asymmetric limit, the instability increments are notably larger than those obtained in the symmetric one. Also, in this limit there is no the drift velocity threshold for the instability, which starts at any $M > 0$ as long as the increment exceeds the collisional damping.

## IV. DISCUSSION

In order for the plasmonic effects that we have discussed to be observed in practice, there are important experimental conditions that should be satisfied. Firstly, we require that $\omega'\tau \gg 1$, where $1/\tau$ is the momentum relaxation rate due to phonon and impurity scattering. In addition, to ensure the coherent growth of the plasma oscillations, the condition $|\omega''| > 1/2\tau$ should also be satisfied [8]. Referring to the data for $M = 0.5$ in Fig. 2(d), we identify $\omega'/2\pi \sim 2.1 \times 10^{12}$ Hz and $\omega'' \sim 10^{12}$ s$^{-1}$. Based on these values, we therefore require $\tau > \sim 0.5$ ps, a value that should be achievable in InGaAs heterostructures at room temperature [13]. We also note that, even for the maximum field condition of $M = 0.5$ considered in Fig. 2(d), the drift velocity $v_0 \approx 3 \times 10^5$ m/s, less than the saturation threshold for this material [29]. Also, at an average mobility $\mu = 2 \times 10^4$ cm$^2$/Vs, the required drain-voltage drop across each plasmonic cavity will be around 30 mV, below the level at which hot-electron effects can be expected to impact transistor performance.

In conclusion, we have proposed a design for a TeraFET with an asymmetric gate-array structure, which should be realizable in conventional III-V-based heterostructure systems. Central to the design of these devices is the introduction of deliberate asymmetry, which is implemented through the inclusion of a narrow gate that serves to break the symmetry within the basic cell of the plasmonic crystal. We have demonstrated that the DS instability in the current-biased transistor is dramatically amplified in the asymmetric elementary cells, reflecting the coherence of plasma oscillations across the entire crystal. By calculating the energy spectrum of the plasmonic crystal and its instability increments, we analyzed the dependence of the latter on the electron drift velocity and the extent of the structural asymmetry. Our results point to the possibility of exciting steady-state plasma oscillations at room temperature, in arrays whose gate asymmetry should be readily implementable via standard nanolithography (as in Fig. 1(c)). We therefore consider this approach to represent a promising pathway to the generation of THz signals, for applications such as wireless networks-on-chip [1] and beyond-6G systems.

**ACKNOWLEDGMENTS**

GRA is thankful to M. Shur for many stimulating discussions. JPB, AM and SM gratefully acknowledge the support of the Coherent/II-VI Foundation.